\documentclass[12pt]{article} 
\begin{document}
%---------------------------------------------------------------------------
\title{\bf Dirty black holes: Symmetries at stationary non-static horizons}
%---------------------------------------------------------------------------
\author{A J M Medved, Damien Martin, and Matt Visser\\
School of Mathematical and Computing Sciences\\
Victoria University of Wellington\\ 
PO Box 600, Wellington\\ 
New Zealand}
%------------------------------------------------
%------------------------------------------------
%---------------------------------------------------------------------------
\date{5 March 2004; \LaTeX-ed \today}
%---------------------------------------------------------------------------
%---------------------------------------------------------------------------
%---------------------------------------------------------------------------
\maketitle
%---------------------------------------------------------------------------
%---------------------------------------------------------------------------

%---------------------------------------------------------------------------
%---------------------------------------------------------------------------
\begin{abstract}
   
  We establish that the Einstein tensor takes on a highly symmetric
  form near the Killing horizon of any stationary but non-static (and
  non-extremal) black hole spacetime. [This follows up on a recent
  article by the current authors, gr-qc/0402069, which considered
  static black holes.]  Specifically, at any such Killing horizon ---
  irrespective of the horizon geometry --- the Einstein tensor
  block-diagonalizes into ``transverse'' and ``parallel'' blocks, and
  its transverse components are proportional to the transverse metric.
  Our findings are supported by two independent procedures; one based
  on the regularity of the on-horizon geometry and another that
  directly utilizes the elegant nature of a bifurcate Killing horizon.
  It is then argued that geometrical symmetries will severely
  constrain the matter near any Killing horizon.  We also speculate on
  how this may be relevant to certain calculations of the black hole
  entropy.

\bigskip

\centerline{gr-qc/0403026}

\bigskip

\centerline{{\sf damien.martin@mcs.vuw.ac.nz}, 
            {\sf joey.medved@mcs.vuw.ac.nz}}
\centerline{ {\sf matt.visser@mcs.vuw.ac.nz}}
\end{abstract}
%----------------------------------------------------------------------------

%\enlargethispage{250pt}
\clearpage

%--------------------------------------------------
\section{Introduction}
%---------------------------------------------------
\def\tr{\hbox{\rm tr}}
\def\implies{\Rightarrow}
\def\conv{\hbox{\rm conv}}
\def\Re{ {\cal R} }
\def\half{{{1\over2}}}
\def\d{{\mathrm{d}}}
\def\quarter{{1\over4}}

There has been a longstanding belief that Bekenstein's black
hole entropy \cite{BEK,HAW},
\begin{equation}
S_{BH} = {1\over 4}\left[{\rm Horizon\;area\;in\;Planck\;units}\right],
\end{equation}
can be explained through a process of state counting.  In the standard
lore, the microstates in question are presumed to arise at the level
of the (yet-to-be-clarified) quantum theory of gravity \cite{SMO}.
Notably, two prime candidates for the fundamental theory --- string
theory \cite{STR-V} and loop quantum gravity \cite{ASH} --- have both
had some (limited) success at statistically calculating this entropy.

An alternative [but perhaps complementary] viewpoint is that $S_{BH}$
can be attributed to a classically inherited symmetry of the black
hole spacetime~\cite{CAR3}. (This symmetry typically manifests as a
two-dimensional conformal field theory.~\footnote{Two-dimensional
  conformal field theories are nice because the entropy is easily
  calculated by way of the Cardy formula~\cite{CARDY}.})  Such a
notion has its origins in a work by Strominger~\cite{STR}, where it
was shown that the entropy of a three-dimensional black hole
\cite{BTZ} is effectively controlled by symmetries that arise out of
the classical property of diffeomorphism invariance~\cite{BH}.  It has
since been a challenge to generalize this type of scenario to four
(and even arbitrary) dimensional spacetimes.  Furthermore, it would be
preferable, ideally speaking, if the controlling symmetry is directly
associated with the black hole horizon~\cite{CAR3}.  (Conversely, in
the Strominger picture, diffeomorphism invariance gives rise to a
conformal theory that lives at spatial infinity.)

There has indeed been substantial progress in the stated directions;
in particular, the treatments by Carlip~\cite{CAR1,CAR2,CAR4} and
Solodukhin~\cite{SOL}.  Nevertheless, it remains unclear as to what
is, precisely, the classical symmetry that underlies these
calculations.  In this regard, Carlip has suggested an ``asymptotic''
conformal field theory in the neighborhood of the horizon \cite{CAR4}.
(Asymptotic in the sense that the symmetry becomes exact only in the
limit as the horizon is approached.)

Assuming Carlip's notion to be essentially correct (for caveats,
see~\cite{MIP}), the current authors have proposed a fundamental
explanation for this asymptotic symmetry~\cite{MMV}.  To elaborate, we
have demonstrated, by purely geometrical arguments, that the Einstein
tensor for {\em any} static black hole spacetime takes on a highly
symmetrical form {\em at the horizon}.~\footnote{This discussion
  applies, strictly speaking, only to a non-extremal horizon.  Unless
  specified otherwise, the assumption of non-extremality will always
  be in effect.}  More specifically, the Einstein tensor
block-diagonalizes into ``transverse'' and ``parallel'' blocks, and
its transverse components are directly proportional to the (induced)
transverse metric.~\footnote{Here and throughout, directional
  terminology --- such as transverse ($\perp$) and parallel
  ($\parallel$) --- refers to the orientation of a (typically
  arbitrary) spacelike cross-section of the horizon.}  The Einstein
field equation then implies that the stress energy tensor on the
static Killing horizon takes the following form:
\begin{equation}
\left.T^{\mu}{}_{\nu}\right|_H  = \left[ \begin{array}{cc|c}
-\rho_H &0 &0\\
0& -\rho_H & 0\\
\hline
0&  0 & \left[T^{\mu}{}_{\nu}\right]_{\parallel} 
\vphantom{\bigg{|}} \end{array} 
\right] \;,
\label{stress}
\end{equation}
where $\rho_H$ is the energy density at the horizon and, by virtue of
symmetry, $[p_\perp]_H=-\rho_H$ is the transverse component of the
pressure.~\footnote{This basic result has been known for some time for
  a static spherically symmetric Killing horizon \cite{DBH1,PAD}. The
  novelty of our recent work~\cite{MMV} was to lift the stringent
  condition of spherical symmetry. For instance, an asymmetric
  ``ring'' of material placed around the equator of a
  Schwarzschild-like black hole would distort the horizon from a
  spherical into an ovaloid shape.}  In particular, we would expect,
on the basis of Ehrenfest's theorem, that any contribution to the
stress energy which arises from averaged quantum effects should take
this form.

The purpose of the current paper is to investigate if these symmetries
persist for a stationary but non-static (black hole) Killing horizon.
As before, the horizon geometry is allowed to be completely general;
that is, we consider ``dirty'' black holes~\cite{DBH1,DBH2,DBH3}. We
are able to establish these on-horizon symmetries by purely
geometrical arguments; thus extending the applicable realm to include
all time-reversible stationary black holes. Moreover, it can be
anticipated, on physical grounds, that a dynamical black hole whose
evolution rate is small [in relation to its surface gravity] will not
lead to a significantly different stress-energy tensor than that
allowed by equation (\ref{stress}).

The rest of the manuscript is organized as follows: In the next
section, we start by invoking some well-known properties of stationary
Killing horizons, with our main focus being on the non-static class.
This allows us to constrain the black hole spacetime metric in the
neighborhood of any such horizon.  (These conditions are in the form
of Taylor-series expansions with respect to the normal distance from
the horizon.) We are able to substantiate these constraints, in
Section 3, by appealing to the regularity of curvature invariants on
the horizon.  Then, in Section 4, after outlining the relevant
methodology, we finally present the on-horizon form of the Einstein
tensor.  Section 5 provides an alternative argument as to why {\em
  any} [bifurcate] stationary Killing horizon must have a Einstein
tensor with the advertised symmetries.  We conclude in Section 6 with
a discussion.

%-------------------------------------------------------------------
\section{Boundary conditions at the horizon}
%-------------------------------------------------------------------

First of all, the following observation will allow us to greatly
simplify the problem at hand: Stationary black holes are expected to
be either static or axially symmetric. (See, for instance, page 523
of~\cite{Wald}.)  To better understand this notion, let us consider
some black hole that is rotating.  If such a black hole is embedded in
a spacetime that is {\em not} axially symmetric, there will be a tidal
frictional force that acts to both slow down the rotation and smooth
out the axial asymmetry.  Consequently, such a spacetime cannot be
stationary; rather, it must still be evolving. However, once the
evolution has finally stopped, there can only be two possibilities:
\begin{itemize}
\item The rotation has completely stopped, in which case the black
  hole is static but not necessarily axially symmetric.
\item The black hole is still rotating, but all of the axial
  asymmetries have been smoothed out so that there is no longer any
  tidal friction. In this case, the black hole is stationary,
  non-static \emph{and axially symmetric}.
\end{itemize}
Considering that the static case has already been covered in our prior
work~\cite{MMV}, we will now  assume an axially symmetric
spacetime (until the more general analysis of Section 5).

Since the spacetime in question has been deemed as axially
symmetric, there must be a rotation axis that picks out a particular
spacelike direction --- say $\phi$ --- and two independent Killing
vectors: a timelike one, $\psi^{\mu}=[\partial_t]^{\mu}\;$, and an
axial one, $\xi^{\mu}=[\partial_{\phi}]^{\mu}\;$.  It is, therefore,
possible to introduce coordinates $(t,\phi,x^2,x^3)$ such that
\begin{equation}
g_{\mu\nu} = \left[ \begin{array}{cccc}
g_{tt}&g_{\phi t} & g_{t2}& g_{t3}\\
g_{\phi t} & g_{\phi\phi}&g_{\phi 2}&g_{\phi 3}\\
g_{t2}&g_{\phi2}&g_{22}&g_{23}\\
g_{t3}&g_{\phi3}&g_{23}&g_{33}
\end{array}\right] \;,
\end{equation}
where the metric components are functions of the remaining spacelike
coordinates, $x^2$ and $x^3$, only.

Let us now impose the additional constraint that our spacetime is
invariant under ``time-reversal'', which means the spacetime is
invariant under the simultaneous change of
\begin{equation}
t \to -t \qquad \hbox{and} \qquad \phi \to -\phi \;.
\end{equation}
(That is, a change in the direction of time should
correspond to a reverse in  the sense of rotation.)
As a consequence, the above metric can
be simplified as follows ({\it cf}, Section 7.1 of~\cite{Wald}): 
\begin{equation}
g_{\mu\nu} = \left[ \begin{array}{cccc}
g_{tt}&g_{\phi t} & 0 & 0\\
g_{\phi t} & g_{\phi\phi}&0&0\\
0&0&g_{22}&g_{23}\\
0&0&g_{23}&g_{33}
\end{array}\right] \;.
\end{equation}
The existence of such a symmetry is, under certain circumstances, a
\emph{theorem} that can be derived from integrability conditions
placed on the Killing vectors. However, we feel that most physicists
would be happy to simply assume the symmetry on physical grounds.

It is convenient to transform the 2$\times$2--block in the
$t$--$\phi$-plane  into an ADM-like form; namely,
\begin{equation}
g_{\mu\nu} = \left[ \begin{array}{cccc}
- [N^2 -g_{\phi t}^2/g_{\phi\phi}] &g_{\phi t} & 0 & 0\\
g_{\phi t} & g_{\phi\phi}&0&0\\
0&0&g_{22}&g_{23}\\
0&0&g_{23}&g_{33}
\end{array}\right]
\;, 
\end{equation}
where $N$ denotes the usual ``lapse'' function.  This formulation
makes it clear that the {\em ergosurface} of the black hole is located
at
\begin{equation}
g_{tt}=0  \iff N^2 = g_{\phi t}^2/g_{\phi\phi} \;,
\end{equation}
while the horizon (or surface of infinite red-shift) is at
\begin{equation}
N = 0 \;.
\end{equation}

To proceed, it proves useful if the 2$\times$2--block in the
$x^2$--$x^3$-plane is appropriately simplified.  Let us call these two
coordinates $n$ and $z$; with $n$ representing the normal distance to
the horizon (such that $N=0$ at $n=0$) and the $z$-direction being
perpendicular to that of $n$. Then, without any loss of generality, we
have
\begin{equation}
g_{\mu\nu} = \left[ \begin{array}{cccc}
- [N^2 -g_{\phi t}^2/g_{\phi\phi}] &g_{\phi t} & 0 & 0\\
g_{\phi t} & g_{\phi\phi}&0&0\\
0&0&1&0\\
0&0&0&g_{zz}
\end{array}\right] \;,
\end{equation}
where $N$, $g_{\phi\phi}$, $g_{\phi t}$ and $g_{z z}$ can now be
regarded as functions of $n$ and $z$.  Such Gaussian normal
coordinates can only be expected to have validity for some region
around the $n=0$ surface, insofar as the geodesics defining our
coordinate system will typically intersect. This will, however, not be
an issue in the analysis, as we are always looking at the near-horizon
limit.

It is also convenient to define an ``angular-rotation'' parameter such
that $\omega \equiv -g_{\phi t}/g_{\phi\phi}\;$, and so
\begin{equation}
g_{\mu\nu} = \left[ \begin{array}{cccc}
- [N^2 -g_{\phi\phi}\;\omega^2] &-g_{\phi\phi} \;\omega& 0 & 0\\
-g_{\phi\phi}\;\omega & g_{\phi\phi}&0&0\\
0&0&1&0\\
0&0&0&g_{zz}
\end{array}\right]
\end{equation}
or, equivalently,
\begin{equation}
\d s^2 = - N(n,z)^2\;\d t^2 + g_{\phi\phi}(n,z) 
\{ \d\phi - \omega(n,z) \;\d t \}^2 
+ \d n^2 + g_{zz}(n,z) \; \d z^2
\;.
\label{x11}
\end{equation}

We can  be even more specific about  the  metric by considering some
well-known properties of stationary Killing horizons.  Firstly, there
is a {\em zeroth law} of black hole mechanics~\cite{Wald,Racz}, which
tells us that the surface gravity or
\begin{equation}
\kappa_H \equiv \lim_{n\to 0}\partial_{n}N 
\label{SG}
\end{equation}
must be a non-negative constant on the horizon.~\footnote{Following an
  analysis which is very similar to that of the appendix in
  \cite{MMV}, one can readily verify that equation (\ref{SG}) complies
  with the standard definition of the surface gravity~\cite{Wald}.}
Note that current considerations will be restricted to non-extremal
horizons, for which $\kappa_H>0$.~\footnote{For the static case, we
  have previously shown that an extremal horizon ($\kappa_H=0$) must
  be located at an infinite proper distance, $n=-\infty$~\cite{MMV}.
  As briefly discussed in Section 3, similar arguments lead to the
  same conclusion for any stationary black hole.}  Secondly, there is
a {\em rigidity theorem} for axially symmetric (stationary, non-static)
Killing horizons~\cite{Carter}, which states that
\begin{equation}
\Omega_H \equiv
-\lim_{n\to 0} {g_{\phi t}\over g_{\phi\phi}} =
\lim_{n\to 0} \omega 
\end{equation}
is also a constant on the horizon.  Finally, let us take note of the
following observation made by Carter [see ~\cite{Carter}, equation
(6.95)]: A stationary Killing horizon is a geodesic submanifold; so
that, if one starts at any point in the horizon and follows the
geodesic along any tangent vector to the horizon, the resultant curve
must then remain in the horizon.  This implies that the horizon is
\emph{extrinsically flat} [see~\cite{Carter}, equation (6.124)];
meaning that the extrinsic curvature of the horizon, itself, must be
zero.  Hence,~\footnote{Note that we use Misner--Thorne--Wheeler
  conventions for the extrinsic curvature. In particular, see page 552
  of~\cite{MTW}.}
\begin{equation}
\lim_{n\to 0} K_{\mu\nu} = - {1\over2} 
\lim_{n\to 0} {\partial g_{\mu\nu}\over\partial n} = 0 \;.
\end{equation}

The above properties directly imply a set of 
necessary constraints, which can be expressed in terms  of Taylor-series
expansions as follows:
\begin{equation}
g_{\phi\phi}(n,z) =  [g_H]_{\phi\phi}(z) + 
{1\over 2} [g_2]_{\phi\phi}(z) \;n^2 + o(n^3)
\;,
\label{x15}
\end{equation}
\begin{equation}
g_{zz}(n,z) =  [g_H]_{zz}(z) + 
{1\over 2} [g_2]_{zz}(z) \; n^2 + o(n^3)
\;,
\label{x16}
\end{equation}
\begin{equation}
\omega(n,z) = \Omega_H + \omega_1(z) \; n 
+ {1\over2} \omega_2(z) \; n^2 +  o(n^3) \; ,
\label{x17}
\end{equation}
and
\begin{equation}
N(n,z)= \kappa_H \; n +  {1\over 2} \kappa_1(z) \; n^2
+ {1\over3!} \kappa_2(z) \; n^3 +  o(n^4)
\;.
\label{x18}
\end{equation}

In analogy with the static case, one might expect to be able to
derive the stronger result~\cite{MMV}
\begin{equation}
N(n,z)= \kappa_H \; n + {1\over3!} \kappa_2(z) \; n^3 +  o(n^4)
\; ,
\label{x19}
\end{equation}
so as to avoid a curvature singularity on the horizon. Although true,
we cannot conclude this from the current argument since $\kappa_1(z)$
could still be a function of $\omega$ that vanishes as
$\omega\rightarrow 0$.  In the next section, we explicitly write out
the Taylor series and demonstrate that equation (\ref{x19}) is indeed
necessary.  Similarly, it is also shown that the on-horizon extrinsic
curvature and the coefficient $\omega_1(z)$ are required to vanish.

%------------------------------------------------------------------
\section{Boundary conditions revisited}
%------------------------------------------------------------------

Next, we will be able to explicitly verify the constraint equations
(\ref{x15}) and (\ref{x16}), strengthen (\ref{x17}) and substantiate
(\ref{x19}) by requiring on-horizon regularity.  More precisely, we
will require a well-behaved horizon 2-geometry and the absence of
curvature singularities on the horizon.  Note that any of the
following calculations can be obtained by a somewhat tedious hand
calculation, although we have, at times, opted for a symbolic
computation using {\it Maple}.

We begin here by recalling the metric of equation (\ref{x11}), as
appropriate for the near-horizon geometry of a stationary Killing
horizon with axial (and time-reversal) symmetry. Keep in mind that the
horizon is located, by construction, at the surface where $n=0$.
(Note that this formalism can, strictly speaking, only be valid for a
non-extremal horizon, as an extremal horizon cannot occur at a finite
value of $n$~\cite{EXT}.  Later on, when we look explicitly at
extremal horizons, this well-known fact will drop out quite
naturally.)

For calculational purposes, let us expand the lapse and the rotation
parameter in the most general manner possible [subject only to the
constraint that the lapse vanishes on the horizon]; that is,
\begin{equation}
N(n,z) = \kappa_H(z) \; n + \frac{1}{2} \kappa_1(z) \; n^2 
+ \frac{1}{3!}\kappa_2(z) \; n^3 + O(n^4) \;,
\end{equation}
\begin{equation}
\omega(n,z) = \omega_H(z) + \omega_1(z) n\;  
+ \frac{1}{2}\omega_2(z)\; n^2 + \frac{1}{3!}\omega_3(z) \; n^3 + O(n^4) \;.
\end{equation}
The following point should be emphasized: We are not assuming our
refined form for the lapse [{\it cf}, equation  (\ref{x19})]
but, rather, attempting to verify  this result  (along with  the other
refinements) by independent means.

Essentially, we are interested in the regularity (or lack thereof) of
the following curvature invariants at the horizon:
\begin{itemize}
\item The Ricci scalar, $R$.
\item The traceless part of the Ricci tensor squared, 
$R_{\mu\nu}R^{\mu\nu} - \frac{1}{4}R^2\;$.
\item The Weyl tensor squared.
\end{itemize}
Clearly, if any of these three scalars are infinite at the horizon,
then a curvature singularity exists.  

First, a calculation of the Ricci scalar yields
\begin{equation}
R= 
\frac{ [g_H]_{\phi\phi}(z) }{2 \; [g_H]_{zz}(z)\; \kappa_H(z)^2}
\left\{[g_H]_{zz}\; \omega_1(z)^2 
+ \left(\frac{\d \omega_H(z)}{\d z}\right)^2\right\} {1\over n^2}
+ o\left(\frac{1}{n}\right) \;.
\label{R}
\end{equation}
[Note that, since the horizon 2-geometry is assumed to be regular,
$g_{zz}$ must be positive on the horizon.]  Now, to avoid a curvature
singularity, the coefficient of any negative power of $n$ must be
zero.  Given that the leading-order term is a sum of squares, we
immediately obtain \emph{two} conditions:
\begin{eqnarray} 
\frac{\d\omega_H(z)}{\d z} &=& 0 \quad 
\Rightarrow \quad\omega_H(z)\; =\; \omega_H\; = \;{\rm constant}\;,\\
\omega_1(z) &=& 0 \;.
\end{eqnarray} 
Hence, we have recovered both the {\em rigidity theorem}
~\cite{Carter} and the anticipated vanishing of the linear-order term.

Second, let us consider the traceless part of the Ricci tensor
squared. Using the previous results to simplify matters, we find that
\begin{eqnarray}
4R_{\mu\nu}R^{\mu\nu} - R^2 &=& 
\Bigg\{ \left(\frac{\d \ln g_{zz}}{\d z}\Big\vert_{n=0}\right)^2 
+ \left(\frac{\d \ln g_{\phi\phi}}{\d z}\Big\vert_{n=0}\right)^2 
\label{RR} \\
&&
+ \frac{16 \; \kappa_1(z)^2}{\kappa_H(z)^2} 
+ \frac{8\left({\d \kappa_H(z)}/{\d z}\right)^2 }
{[g_H]_{zz}(z) \;\kappa_H(z)}\Bigg\} \; {1\over n^2}
+ o\left(\frac{1}{n}\right) \;. \nonumber 
\end{eqnarray}
This is, once again, a sum of squares, from which we can deduce the
following {\em four} constraints:
\begin{eqnarray}
\frac{\d\kappa_H(z)}{\d z} &=& 0 \quad 
\Rightarrow \quad
\kappa_H(z) \;=\; \kappa_H \;=\;{\rm constant}  \;,\\
\kappa_1(z) &=& 0 \;, \\
\frac{\d g_{zz}}{\d z}\vert_{n=0} 
&=& 0 \quad \Rightarrow \quad g_{zz}\;=\;[g_{zz}(z)]_H \;+\;o(n^2) \;,\\
 \frac{\d g_{\phi\phi}}{\d z}\vert_{n=0} &=& 0 
\quad\Rightarrow\quad g_{\phi\phi}\;=\; [g_{\phi\phi}(z)]_H \;+\; o(n^2)\;.
\end{eqnarray}
Note that the first of these conditions allows us to recover the {\em
  zeroth law}~\cite{Wald,Racz}.

Finally, after imposing all the above constraints, we find that the
Weyl tensor squared yields a finite result as $n\to 0$ [{\it i.e.},
the leading order is at least $o(1)$].  Therefore, our regularity
requirements leave us with a set of {\emph{six}}
constraints;~\footnote{One can use the above conditions to verify that
  no further constraints can be extracted from the $o(n^{-1})$ terms
  in equations (\ref{R}) and (\ref{RR}).}  all of which are necessary
and, in combination,  sufficient for the horizon not to have any
curvature singularities.  In this unambiguous way, we have
substantiated the series expansions in equations
(\ref{x15})--(\ref{x17}) [with $\omega_1=0$] and equation (\ref{x19}).
Moreover, the above statements can be regarded as a rigorous proof,
which supersedes our earlier plausibility arguments.

Let us now address the issue of extremal horizons, for which
$\kappa_H=0$.  Assuming an $m$-order degeneracy in the surface
gravity, we can expand the lapse as
\begin{equation}
N(n,z) = \frac{\kappa_m(z)}{m!} \; n^m 
+ \frac{\kappa_{m+1}(z)}{(m+1)!}\; n^{m+1} + o(n^{m+2}) \;.
\end{equation}
Now proceeding just as before, we find that, 
\begin{equation}
R = \frac{[g_H]_{\phi\phi} \; m! \; 
\left\{[g_H]_{zz} \; \omega_1(z)^2 
+ \left({\d \omega_0(z)}/{\d z}\right)^2\right\} }
{2\; [g_H]_{zz}\; \kappa_m(z)^2} \;
n^{-2m} + o(n^{-2m+1}) \;,
\end{equation}
which informs us that the \emph{rigidity theorem} still holds and 
the linear term $\omega_1$ vanishes. Working through the subdominant
terms, we can  place still more constraints on $\omega(n,z)$. 
Then, after some
effort, we arrive at
\begin{equation}
R_{\mu\nu} \; R^{\mu\nu} = 
\frac{m^2(m-1)^2}{n^4} + 
\frac{1}{4} \; 
\frac{ [g_H]_{\phi\phi}^2 \; m^4}
{n^4 \; [g_H]_{zz}^2 \;\kappa_m(z)^4}
\left(\frac{\d \omega_{m-1}(z)}{\d z}\right)^4 
+ o\left({1\over n^3}\right) \;, 
\end{equation}
which is a sum of squares.  Because the first term is required to vanish,
either $m=1$ (a non-extremal horizon at $n=0$) or $m=0$ (no
horizon at $n=0$).  Since any finite value of $n$ can always be
shifted to $n=0$, this really tells us that an extremal horizon cannot
occur at a finite value of the normal coordinate. Hence, as in the
static case, any extremal horizon must be located at $n=-\infty$.

%-------------------------------------------------------------------
\section{The on-horizon Einstein tensor}
%-------------------------------------------------------------------

We can now, with the help of a symbolic {\em Maple} computation, use
these Taylor-series expansions to deduce the various components of the
on-horizon Einstein tensor.  However, to express this tensor in a form
that emphasizes its symmetrical nature, it will first be necessary to
introduce some additional formalism.

To begin, let us take note of the null Killing vector for a stationary
(black hole) Killing horizon. That is (for instance, \cite{Wald}),
\begin{equation}
\chi^a = \xi^a + \Omega_H \; \psi^a \;.
\end{equation}
Then, in terms of our $(t,\phi,n,z)$ coordinate system,
we have
\begin{equation}
\chi^a = (1,\Omega_H,0,0)
\;,
\end{equation}
\begin{equation}
\xi^a = (1,0,0,0)\;,
\end{equation}
and
\begin{equation}
\psi^a = (0,1,0,0)
\;.
\end{equation}

Let us next consider the contraction
\begin{equation}
g(\chi,\chi) \equiv g_{ab} \; \chi^a \; \chi^b 
= g_{tt} + 2\Omega_H \; g_{t\phi} + \Omega_H^2 \; g_{\phi\phi} =
- N^2 + g_{\phi\phi} \; [\Omega_H-\omega]^2 \;,
\end{equation}
which, by an inspection of  our Taylor-series expansions,  implies that
\begin{equation}
g(\chi,\chi) = - \kappa_H^2 \; n^2 +   o(n^4)
\;.
\label{x37}
\end{equation}
It would then seem sensible to define, \emph{outside of  the event horizon},
a normalized vector of the form
\begin{equation}
\hat \chi = {\chi \over\sqrt{-g(\chi,\chi)}}
\;.
\end{equation}

We can similarly write 
\begin{equation}
g(\psi,\psi) = g_{\phi\phi} = o(1)
\end{equation}
and 
\begin{equation}
\hat \psi = {\psi \over\sqrt{g(\psi,\psi)}} = {\psi \over\sqrt{g_{\phi\phi}}}
\;.
\end{equation}
Furthermore, 
\begin{equation}
g(\chi,\psi) = g_{ab} \; \chi^a \; \psi^b = g_{t\psi}+ \Omega_H \; g_{\psi\psi}
= g_{\psi\psi} \; (\Omega_H-\omega) = o(n^2)
\end{equation}
and 
\begin{equation}
g(\hat\chi,\hat\psi) = o(n) \;.
\end{equation}
Finally, 
\begin{equation}
\hat n = n =  (0,0,1,0)
\end{equation}
and
\begin{equation} 
\hat z = {(0,0,0,1)\over\sqrt{g_{zz}}}
\;.
\end{equation}

Granted, $\hat \chi$ and $\hat \psi$ are not exactly orthonormal.
But these vectors are certainly orthonormal to the order $o(n)$ and span the
space perpendicular to the horizon, which is enough to investigate the
on-horizon geometry.

We are now suitably positioned to discuss the on-horizon components of
the Einstein tensor. For this purpose, it is helpful to define
\begin{eqnarray}
[G_{\hat\chi\hat\chi}]_H &\equiv& \lim_{n\to0} G(\hat \chi,\hat\chi) 
= \lim_{n\to0} {G(\chi,\chi)\over g(\chi,\chi)} 
\nonumber
\\
&& = 
\lim_{n\to0} {G_{tt} + 2\Omega_H \; G_{t\phi} 
+\Omega_H^2 \; G_{\phi\phi} \over \kappa_H^2 \; n^2}\;,
\\{}
[G_{nn}]_H & \equiv & \lim_{n\to0} G(n, n)  = \lim_{n\to0} G_{nn} 
\;,
\\{}
[G_{\hat z\hat z}]_H &\equiv& \lim_{n\to0} G(\hat z,\hat z)  =
\lim_{n\to0} {G_{zz}\over g_{zz}}
\;,
\\{}
\left[ G_{\hat\phi\hat\phi}\right]_H &\equiv& 
\lim_{n\to0} G(\hat \psi,\hat \psi)  =
\lim_{n\to0} {G(\psi,\psi)\over g(\psi,\psi)} =
\lim_{n\to0} {G_{\phi\phi}\over g_{\phi\phi}}
\;,
\\{}
[G_{\hat\phi\hat\chi}]_H
&\equiv& \lim_{n\to0} G(\hat \chi,\hat \psi) 
= 
\lim_{n\to0} {G(\chi,\psi)\over \sqrt{ g(\chi,\chi) \; g(\psi,\psi) } }
\nonumber
\\
&&
=
\lim_{n\to0} {G_{t\phi}+\Omega_H \; G_{\phi\phi} 
\over \kappa_H n \sqrt{ g(\psi,\psi) } }
\;,
\\{}
[G_{n\hat z}]_H&\equiv& \lim_{n\to0} G(n,\hat z) = 
\lim_{n\to0} {G_{nz}\over \sqrt{ g_{zz} } }\;,
\end{eqnarray}
with all other components automatically vanishing by virtue of the
axial symmetry and/or time-reversal symmetry.

Ideally speaking, what we would now like to prove is the following:
\begin{equation}
[G_{\hat\chi\hat\chi}]_H = - [G_{nn}]_H
\;, \label{key1}
\end{equation}
\begin{equation}
 [G_{\hat\phi\hat\chi}]_H  =  0 
\label{key2}
\end{equation}
and
\begin{equation}
 [G_{n\hat z}]_H =  0
\;. \label{key3} 
\end{equation}
Such an outcome would confirm that, just as for a static Killing
horizon~\cite{MMV}, the Einstein tensor block-diagonalizes and the
transverse components of this tensor are proportional to the
transverse metric.  (By transverse, we mean the components orthogonal
to any spacelike cross-section of the event horizon.)  Before
elaborating further on our results, let us point out that there is, in
fact, a strong physical motivation for believing in the vanishing of
$[G_{\hat\phi\hat\chi}]_H$. By the Einstein equations, this term is
equivalent to a non-zero angular ``flux'' at the horizon.
Significantly, the null Killing vector, $\chi$, effectively ``rotates
with the horizon''.  Hence, if any such flux does exist, the
implication would be that the dirt surrounding the black hole is
``moving with respect to the horizon''.  But this would then torque
the black hole --- either spinning it up or slowing it down --- until
the true state of stationarity is finally achieved.  (Alas, we have no
analogously simple  argument for the vanishing of the ``stress'' term
$[G_{n\hat z}]_H$.)

All three of these symmetries have indeed been verified by a symbolic
computation.  To briefly elaborate, using our Taylor-series expansions
[{\it i.e.}, equations (\ref{x15})--(\ref{x17}) with $\omega_1=0$ 
and (\ref{x19})], we
``ask'' {\em Maple} to calculate the Einstein tensor and, afterwards,
take the $n\rightarrow 0$ limit.  A simple inspection then confirms
the validity of equations (\ref{key1})--(\ref{key3}).

For the sake of completeness, we will also present the explicit form
of these on-horizon tensor components.  It is useful, however, if we
first introduce a few more relevant expressions.  For instance,
considering just the in-horizon 2-metric or
\begin{equation}
\d s^2_{\parallel} = g_{\phi\phi}(z)\;  \d\phi^2 +  g_{zz}(z) \; \d z^2 \;,
\label{para}
\end{equation}
we can calculate the corresponding scalar curvature and obtain
\begin{equation}
R_\parallel = \half \left\{ 
 {(\partial_z [g_H]_{\phi\phi})^2\over [g_H]_{zz}\;[g_H]_{\phi\phi}^2}
+ {\partial_z [g_H]_{\phi\phi} \; \partial_z [g_H]_{zz}\over 
[g_H]_{zz}^2\; [g_H]_{\phi\phi}}
- 2{\partial_z^2 [g_H]_{\phi\phi}\over [g_H]_{zz}\;[g_H]_{\phi\phi}}
\right\}
\;.
\end{equation}
Similarly, focusing on the $\hat\chi$--$n$-plane and looking at
the induced transverse 2-metric,
\begin{equation}
\d s^2_\perp = 
-\left[N^2 -g_{\phi\phi} \; (\omega - \Omega_H)^2\right]\d \tilde\chi^2
+ \d n^2
\;,
\end{equation}
we find a corresponding Ricci scalar of the simple form
\begin{equation}
R_\perp =-2  \left\{ {\kappa_2\over\kappa_H } \right\} 
+ \frac{3}{2} \; \frac{[g_H]_{\phi\phi}(z) \; \omega_2(z)^2}{\kappa_H^2}
\;.
\end{equation}
It should be noted that, in obtaining this last result, we had to
expand $g(\chi,\chi)$ out to the \emph{fourth} order in $n$ [so that
equation (\ref{x37}) is insufficient], as non-trivial contributions
occur at this order as a consequence of the normal derivatives.

In terms of the above formalism, the non-vanishing components of the
on-horizon Einstein tensor are now expressible as follows:
\begin{eqnarray}
[G_{\hat\chi\hat\chi}]_H &=& - \half R_\parallel -
\half \tr [g_2]
-  {1\over 4} [g_H]_{\phi\phi} \; {\omega_2^2\over \kappa_H^2}
\;,
\\{}
[G_{nn}]_H &=&- [G_{\hat\chi\hat\chi}]_H
\;,
\\{}
[G_{\hat z\hat z}]_H &=&  - \half R_\perp
+ 
{\left[ g_2\right]_{\phi\phi}\over \left[ g_H\right]_{\phi\phi}}
+ {1\over 2} \; \left[ g_H\right]_{\phi\phi} \; {\omega_2^2\over \kappa_H^2}
\;,
\\{}
[G_{\hat \phi\hat \phi}]_H &=& - \half R_\perp  +
{[g_2]_{zz}\over [g_H]_{zz}}\;,
\end{eqnarray}
where the trace operation (depicted by $\tr$ in the first equation)
has been performed with the in-horizon 2-metric defined by equation
(\ref{para}).

It is easily confirmed that the above results correctly limit to their
static analogues ({\it cf},~\cite{MMV}) as $\omega\rightarrow 0$.  In
spite of the obvious complexities that arise for a stationary but
non-static horizon, the current case is still, in some sense, a
simplification from the most general static formalism.  This is
because the property of axial symmetry now implies that $g_\parallel$
(and, hence, $[g_2]_{\parallel}$ in particular) is diagonal in the
$\phi$--$z$-coordinates. Thus, $G_\parallel$ is automatically
diagonal, which was not necessarily true in the static case.  On the
other hand, there are now extra contributions from $\omega_2$.  Note
that $\Omega_H$ drops out of the on-horizon Einstein tensor completely
--- this can be viewed as a side effect of having a rigidly rotating
horizon.

%-------------------------------------------------------------------
\section{An alternative method: Bifurcate Killing horizons}
%-------------------------------------------------------------------

It has now been confirmed
that the anticipated symmetries in the Einstein
tensor [namely, equations (\ref{key1})-(\ref{key3})] are valid at any
stationary (non-extremal) Killing horizon; but  what is still  lacking is
a clear physical motivation for this phenomenon.  Here, we will help
fill this gap by providing a relatively simple and physically
compelling argument for this highly symmetric form.  Although the
upcoming analysis applies irrespective of axial symmetry, there is one
important caveat: We will now restrict considerations to bifurcate
Killing horizons. Which is to say, it will now be assumed that the
horizon contains a bifurcation surface --- that is, a cross-sectional
(spacelike) 2-surface where the null Killing vector, $\chi^{a}$, is
precisely vanishing.  At a first glance, this may appear to be a
highly restrictive constraint on the spacetime; in particular, a physical
black hole that forms from stellar collapse will not be of this type.
Nonetheless, a physical black hole will asymptotically approach such a
spacetime. Indeed, Racz and Wald have shown that, if the surface
gravity is constant and non-vanishing over a patch of Killing horizon
(containing a spacelike cross-section), there will exist a stationary
extension of the spacetime which does include a regular bifurcation
surface~\cite{Racz2}.  Since the zeroth law is automatically satisfied
for any stationary Killing horizon~\cite{Racz}, the existence of such
an extension will, for our purposes, always be ensured.

For the analysis of this section, it proves to be convenient if we
employ a different basis for the (on-horizon) coordinate system.  To
set up a suitable basis, let us start by considering an arbitrary spacelike
section of the horizon.  Like all spacelike 2-surfaces, it is possible
--- at every point --- to find two null vectors that are orthogonal to
the section, as well as to each other.  Let us choose one of these to
be the Killing vector $\chi^{a}$ and denote the other by $N^{a}$. For
the sake of convenience, we will adopt the normalization $\chi^a \;N_a
= -1$. (And since both of these are null, $\chi^a \;\chi_a = N^a \;N_a =
0$.)

To complete our coordinate basis, we can choose any pair of orthogonal
spacelike vectors, $m_1^a$ and $m_2^a$, that are tangent to the
horizon section in question.  It should be clear that, by
construction, $\chi^a$, $N^a$, $m_1^a$, and $m_2^a$ form an
orthonormal basis for the tangent space. Consequently, there will
exist coefficients such that the Einstein tensor can be written as
follows:
\begin{eqnarray}
G_{ab} &=& G_{++} \;\chi_a \chi_b +  G_{--} \; N_a N_b 
+ G_{+-} \;\left\{ \chi_a N_b + N_a \chi_b \right\}
\nonumber \\ 
&&
+ G_{+1} \;\left\{ \chi_a [m_1]_b + [m_1]_a \chi_b \right\}
+ G_{+2} \;\left\{ \chi_a [m_2]_b + [m_2]_a \chi_b \right\}
\nonumber \\
&&
+ G_{-1} \;\left\{ N_a [m_1]_b + [m_1]_a N_b \right\}
+ G_{-2} \;\left\{ N_a [m_2]_b + [m_2]_a N_b \right\}
\nonumber \\
&&
+ G_{11} \; [m_1]_a [m_1]_b +  G_{22} \; [m_2]_a [m_2]_b
\nonumber \\
&&
+ G_{12}  \; \left\{ [m_1]_a [m_2]_b + [m_2]_a [m_1]_b \right\}
\;.
\label{ET}
\end{eqnarray}
(With an analogous form, in fact, for any symmetric tensor.)

We can significantly simplify the above expression by, firstly, taking
note of equation (7.1.15) from Wald's textbook~\cite{Wald}
(re-expressed in one-form notation),
\begin{equation}
\chi \wedge (R \cdot \chi) = 0 \;,
\end{equation}
which is valid [{\em on the horizon}] for any stationary Killing
horizon. (Here, $R$ represents the two-form Ricci tensor rather than
the scalar curvature.)  It immediately follows that, on the horizon,
\begin{equation}
(R\cdot \chi) \propto \chi
\end{equation}
or, equivalently,
\begin{equation}
R^b{}_a \; \chi_b \propto \chi_a\;.
\end{equation}
But, since $g^b{}_a\; \chi_b=\chi_a\;$, this also means that
\begin{equation}
G^b{}_a \; \chi_b \propto \chi_a\;,
\end{equation}
and so the Einstein tensor (like the Ricci tensor) possesses a null
eigenvector on the horizon.

To make use of this last property, let us contract
 equation (\ref{ET}) with $\chi^b\;$ and then apply the
orthogonality properties. This process yields
\begin{equation}
G_{ab} \;\chi^b = - G_{--} \; N_a - G_{+-} \; \chi_a 
- G_{-1} \; [m_1]_a  - G_{-2} \; [m_2]_a
\;. 
\end{equation}
However, we know that, on the horizon, this contraction must be
proportional to $\chi_a$; thus implying
\begin{equation}
G_{--} = G_{-1} = G_{-2} = 0\;,
\end{equation}
so that the on-horizon  Einstein tensor reduces to
\begin{eqnarray}
G_{ab} &=& G_{++} \; \chi_a \chi_b 
+ G_{+-} \; \left\{ \chi_a N_b + N_a \chi_b \right\}
\nonumber \\
&&
+ G_{+1} \; \left\{ \chi_a [m_1]_b + [m_1]_a \chi_b \right\}
+ G_{+2} \; \left\{ \chi_a [m_2]_b + [m_2]_a \chi_b \right\}
\nonumber \\
&&
+ G_{11} \; [m_1]_a [m_1]_b +  G_{22} \;[m_2]_a [m_2]_b
\nonumber \\
&&
+ G_{12}  \; \left\{ [m_1]_a [m_2]_b + [m_2]_a [m_1]_b \right\} \;.
\end{eqnarray}

This is as far as we can go on an arbitrary section of the horizon, so
let us now specialize to the bifurcation surface.  First note that
\begin{equation}
[g_\perp]_{ab}=
\chi_a N_b + N_a \chi_b
\end{equation}
has a well-defined limit on the bifurcation 2-surface, even though
$\chi^a\to 0$ as the surface is approached. 
This is because the second null normal limits there as 
$N^{a}\to\infty\;$,  since
it still must satisfy $\chi^a \;N_a = -1\;$. It follows that the above
 combination 
simply limits to the 2-metric perpendicular to the bifurcation
surface.  In view of this observation,
the Einstein tensor
 at the bifurcation 2-surface takes on
the greatly simplified form
\begin{eqnarray}
G_{ab} &=& G_{+-} \; [g_\perp]_{ab}
\nonumber \\
&&
+ G_{11} \; [m_1]_a [m_1]_b +  G_{22}\; [m_2]_a [m_2]_b
\nonumber \\
&&
+ G_{12}  \; \left\{ [m_1]_a [m_2]_b + [m_2]_a [m_1]_b \right\} 
\end{eqnarray}
or, in perhaps  more sensible notation,
\begin{eqnarray}
G_{ab} &=& G_{+-} \; [g_\perp]_{ab} + [G_\parallel]_{ab} \;.
\end{eqnarray}
Notice that, on the bifurcation surface, $G_{ab}$ indeed
block-diagonalizes and its transverse components are proportional to
the transverse metric.  This is precisely the form of Einstein tensor
we have been setting out to show!  We can now use the fact that the
Einstein tensor is a Killing invariant to Lie propagate this form away
from the bifurcation surface and, then, onto any other spacelike
section of the horizon.  (See, for instance, a relevant discussion
in~\cite{JKM}.)  In this way, we are able to substantiate the highly
symmetric form of the on-horizon Einstein tensor, but with a
completely independent and (perhaps) physically more translucent
method.\footnote{The present analysis, however, does not help in
  computing the actual value of the Einstein tensor.  Such a
  computation would require something like the analysis of the
  previous section.}

%-------------------------------------------------------------------
\section{Discussion}
%-------------------------------------------------------------------

Let us now briefly summarize the findings of this paper (in
conjunction with our prior treatment~\cite{MMV}): We have been able to
establish a very high degree of symmetry in the Einstein tensor
near {\em any} stationary (static or non-static) 
non-extremal Killing horizon.
In particular, the on-horizon form of this tensor
will block-diagonalize, and its transverse components will be directly
proportional to the transverse metric.  As a direct consequence
(assuming the Einstein field equations), the stress tensor near any
such Killing horizon will be highly constrained. Most notably, the sum
of the energy density and the transverse pressure will tend to zero as
the horizon approached. It should be re-emphasized that such
constraints would apply to all forms of matter or energy in the
proximity of the horizon; including any quantum-induced fluctuations.

Let us recall Carlip's proposal that the black hole entropy is
controlled by an asymptotic near-horizon conformal
symmetry~\cite{CAR4}, as this idea served as one of the principal
motivations for our work. At first a glance, it may not be evident how
our symmetries can be responsible for the type of conformal symmetry
that Carlip had in mind. Nevertheless, we would suggest that matter
obeying the near-horizon form $T_\perp\propto g_\perp$ will
effectively behave as a collection of world-sheet (two-dimensional)
conformal field theories; each of which is defined at a point on the
horizon and acts in the transverse plane. As for the parallel block of
the stress-tensor, Carlip argues that the physics relevant to
leading-order calculations of the black hole entropy should probably
be limited to the transverse plane~\cite{CAR3}.

Important future work should include suitable generalizations to truly
dynamical spacetimes. Such generalizations will, however, likely be
limited to spacetimes with a high degree of symmetry due to the
technically difficult nature of the problem.  It should be
re-emphasized that if the black hole is evolving slowly enough --- on
a scale set by the surface gravity --- then the near-horizon geometry
could be viewed as approximately stationary.  (This notion of a
dynamical black hole being a quasi-stationary entity is conceptually
similar to the {\it isolated horizon} framework of Ashtekar {\it et
  al} \cite{Ashtekar}.)  It can thus be expected that any of our
symmetries remain valid up to corrections that go as the evolution
rate.

%------------------------------------------------------------------
\section*{Acknowledgements}
%------------------------------------------------------------------

Research supported by the Marsden Fund administered by the Royal
Society of New Zealand, and by the University Research Fund of
Victoria University.

%----------------------------------------------------------------------

%-----------------------------------------------------------------------

%-----------------------------------------------------------------------
\end{document}